\documentclass{aa}
\usepackage{lineno}
\usepackage{lscape}
\usepackage{lipsum}
\usepackage[varg]{txfonts}
\usepackage{graphicx}
\usepackage[colorlinks=true, citecolor=blue, urlcolor=blue]{hyperref}
\usepackage{placeins}
\usepackage{xcolor}
\usepackage{soul}

\begin{document}

   \title{Implications of multiwavelength spectrum on cosmic-ray acceleration in blazar TXS~0506+056}

   %\subtitle{I. Overviewing the $\kappa$-mechanism}

   \author{Saikat Das
          \inst{1}
          \and
          Nayantara Gupta\inst{2}
          %\fnmsep\thanks{Just to show the usage of the elements in the author field}
          \and
          Soebur Razzaque\inst{3,4,5}
          }

   \institute{
        Center for Gravitational Physics, Yukawa Institute for Theoretical Physics, Kyoto University, Kyoto 606-8502, Japan\\
        \email{saikat.das@yukawa.kyoto-u.ac.jp}
    \and
         Astronomy \& Astrophysics Group, Raman Research Institute, Bangalore 560080, Karnataka, India\\
         \email{nayan@rri.res.in}
    \and
        Centre for Astro-Particle Physics (CAPP) and Department of Physics, University of Johannesburg, PO Box 524, Auckland Park 2006, South Africa\\
        \email{srazzaque@uj.ac.za}
        %\thanks{The university of heaven temporarily does not accept e-mails}
    \and
        Department of Physics, The George Washington University, Washington, DC 20052, USA
    \and
        National Institute for Theoretical and Computational Sciences (NITheCS), South Africa
    }

  \date{Received YYYY; accepted ZZZZ}

% \abstract{}{}{}{}{} 
% 5 {} token are mandatory
 
  \abstract
  % context heading (optional)
  % {} leave it empty if necessary  
   {
   The MAGIC collaboration has recently analyzed data from a long-term multiwavelength campaign of the $\gamma$-ray blazar TXS 0506+056. 
   %\st{which is associated with IceCube neutrino events.} 
   In December 2018, it was flaring in the very-high-energy (VHE; $E>100$ GeV) $\gamma$-ray band, but no simultaneous neutrino event was detected.
   }
  % aims heading (mandatory)
   {
   %\st{We probe any temporal correlation between the MAGIC VHE $\gamma$-ray data and IceCube neutrino signal.} 
   We model the observed spectral energy distribution (SED), using a one-zone leptohadronic emission.
   }
  % methods heading (mandatory)
   {
   %\st{We model multi-messenger signals due to cosmic-ray (CR) acceleration in the energy range 10 GeV -- 100 EeV, in addition to leptonic emission}.
   %\st{We obtain constraints on the neutrino flux by requiring secondary radiation from hadronic cascade to not overshoot the X-ray data.} 
   We estimate the neutrino flux through the restriction from observed X-ray flux on the secondary radiation due to hadronic cascade, initiated by protons with energy $E_p \lesssim 0.1$ EeV. We assume ultrahigh-energy cosmic rays (UHECRs; $E\gtrsim0.1$ EeV), with the same slope and normalization as the low-energy spectrum, are accelerated in the jet but escape efficiently.
   %with a faster escape rate.} 
   We propagate the UHE protons in a random, turbulent extragalactic magnetic field (EGMF). 
   %\st{and estimate the contribution of line-of-sight cosmogenic $\gamma$-rays to the SED.} 
   }
  % results heading (mandatory)
   {
   The leptonic emission from the jet dominates the GeV range, whereas the cascade emission from CR interactions in the jet contributes substantially to the X-ray and VHE range. The line-of-sight cosmogenic $\gamma$ rays from UHECRs produce a hardening in the VHE spectrum. 
   %that can be tested with upcoming telescopes such as the CTA. 
  Our model prediction for neutrinos from the jet is consistent with the 7.5-year flux limit by IceCube and shows no variability during the MAGIC campaign. Therefore, we infer that the correlation between GeV-TeV $\gamma$-rays and neutrino flare is minimal. The luminosity in CRs limits the cosmogenic $\gamma$-ray flux, which, in turn, bounds the RMS value of the EGMF to $\gtrsim 10^{-5}$ nG. The cosmogenic neutrino flux is lower than the IceCube-Gen2 detection potential for 10 yrs of observation. 
   }
  % conclusions heading (optional), leave it empty if necessary 
   {
   VHE $\gamma$-ray variability should arise from increased activity inside the jet; thus, detecting steady flux at multi-TeV energies may indicate UHECR acceleration. Upcoming $\gamma$-ray imaging telescopes, such as the CTA, will be able to constrain the cosmogenic $\gamma$-ray component in the SED of TXS 0506+056.
   }

   \keywords{Astroparticle physics -- 
   	         galaxies: active -- 
             gamma-rays: general -- 
             neutrinos
            }

   \maketitle
%
%-------------------------------------------------------------------

\section{Introduction}

Blazars are a subclass of radio-loud Active Galactic Nuclei (AGNs), with their highly relativistic jet collimated towards the observer's line of sight. They have been considered as prominent candidates for the origin of IceCube-detected diffuse astrophysical neutrino flux beyond $\sim$10 TeV \citep{Aartsen_2013, Eichler_1979, Sikora_1987, Petropoulou:2015upa, Murase:2018iyl, Yuan_2020} and may also contribute in the PeV-EeV energy range  \citep{Kalashev:2013vba, Kochocki:2020iie, Das:2020hev}. For the first time in September 2017, a high-energy muon-neutrino event IC-170922A ($E_\nu \sim0.3$ PeV) was associated with the $\gamma$-ray flaring blazar TXS 0506+056 at 3$\sigma$ significance \citep{IceCube:2018dnn}. Subsequently, other events having positional coincidence with Fermi-LAT detected blazars are also observed with lower statistical significance \citep{Fermi-LAT:2019hte, Franckowiak:2020qrq}. Despite several studies on this object, it is crucial to revisit the spectral properties for predicting the multi-messenger signals from similar sources.

The explanation of the neutrino event requires synchrotron (SYN) and synchrotron self-Compton (SSC) photons as the target for $p\gamma$ interactions \citep{Gao:2018mnu, Cerruti:2018tmc, Sahu:2020eep}. Whereas, other models require an external photon field, resulting in external inverse-Compton (IC) emission \citep{Reimer:2018vvw, Keivani:2018rnh, Rodrigues:2018tku, Petropoulou:2019zqp}. In the hadronuclear interpretation via $pp$ interaction, the shock accelerated protons may interact with gas clouds in the vicinity of the supermassive black hole \citep{Liu:2018utd} or cold protons in the jet \citep{Banik:2019jlm}. Some studies invoke neutrino production from the interaction of relativistic neutron beams in the jet, originating in $p\gamma$ interactions with external photons \citep[see for e.g.,][]{Zhang:2019htg}. In \cite{Fraija:2020zjk}, $p\gamma$ interactions occur with seed photons produced by annihilation of electron-positron pairs from the accretion disk.

The CR-induced cascade from Bethe-Heitler (BH) pair production contributes near the X-ray energies in the $p\gamma$ scenario, thus also limiting the hadronic component in GeV-TeV $\gamma$-rays. As a result, this constrains the astrophysical neutrinos, in many cases, to a flux level lower than predicted by  IceCube observations. Thus, an additional photon field of energy $\epsilon \simeq m_\pi m_p c^4/20E_\nu \simeq$ 440 eV, i.e., in the UV to soft X-ray energy band, is required. However, an ``orphan'' neutrino flare from this source from September 2014 to March 2015 is revealed from the analysis of archival data, at 3.5$\sigma$ statistical significance \citep{IceCube:2018cha} with $13\pm 5$ signal events above the atmospheric background. The latter was not accompanied by increased activity in $\gamma$-rays, indicating different astrophysical processes may dominate for neutrino and $\gamma$-ray flares. Often, a two-zone model is employed, considering a high opacity for GeV $\gamma$-rays in the neutrino production region \citep{Sahakyan:2018voh,  Xue:2019txw, Xue:2020kuw}.

Recently, the MAGIC collaboration has modeled the spectrum of TXS 0506+056, observed during a multiwavelength campaign lasting 16 months from November 2017 to February 2019, covering the radio band, optical/UV, high-energy, and very-high-energy (VHE, $E>100$ GeV) $\gamma$-rays \citep{MAGIC:2022gyf}. A $\gamma$-ray flaring activity was observed by MAGIC during December 2018. Fermi-LAT observed several short flares on timescales of days to weeks, unlike the long-term flare of 2017. At lower energies, no significant variability was observed. The observed flare was not associated with any neutrino event. Their model infers the neutrino luminosity to be lower than the detection threshold of currently operating instruments.

We analyze the multiwavelength SED using a one-zone leptohadronic model. The low-energy peak results from the SYN radiation of relativistic electrons. The high-energy peak is produced by SSC and IC scattering of external photons (external Compton, abbv.\ EC). These external photons may originate from the broad-line region (BLR). Although a broad-line emission in the optical spectrum is not detected, TXS 0506+056 can be a masquerading BL Lac, i.e., intrinsically a flat spectrum radio quasar (FSRQ) with hidden broad lines and a standard accretion disk \citep{Padovani:2019xcv}. The interaction of the cosmic ray protons with the leptonic radiation and the external photon field produces a characteristic photon spectrum resulting from the electromagnetic cascade of secondary electrons, which is constrained by the X-ray data. 

Blazars are also suitable candidates for ultrahigh-energy cosmic ray (UHECR; $E\gtrsim10^{18}$ eV) acceleration. They can escape the 
%blazar 
jet and interact with cosmic background photons. In an earlier work, the neutrino flux originating in extragalactic propagation of UHECRs from TXS 0506+056, and a few other blazars was analyzed \citep{Das:2021cdf}. We assumed a correlation between the cosmic-ray and IceCube-detected neutrino luminosity inside the jet, to scale the cosmogenic fluxes. Here, we constrain the UHECR luminosity by SED modeling, consistently with the allowed flux of cosmogenic $\gamma$-rays. The EGMF is crucial to determine this line-of-sight resolved $\gamma$-ray component. Using the latest spectrum data, we coherently explain the multiwavelength SED, and predict the corresponding neutrino flux from the jet emission region, and the plausibility of cosmogenic $\gamma$-ray contribution to the SED. Finally, the luminosity constraint can be used to predict the resulting cosmogenic neutrino flux at EeV energies.

The observed multiwavelength SED is difficult to resolve into components coming from interactions inside the jet or line-of-sight resolved UHECR interactions, or multiple zones, etc. Hence all photons are treated on equal footing by the detectors. 
Thus, if UHECRs escape from the source, they produce cosmogenic gamma-rays and the line-of-sight resolved component of that flux can contribute at the VHE range of the MWL SED. We invoke three components (i) a purely leptonic emission (ii) hadronic emission inside the jet (iii) a line-of-sight component of the hadronic emission during extragalactic propagation. This is the reason the cosmogenic gamma-ray spectrum is also used to fit the observed SED.

Our study is essentially a one-zone model, with all the jet parameters constrained by the MWL SED. These jet parameters are used to calculate the luminosity in cosmic rays required inside the jet for SED and neutrino modeling. However, the parameter space is degenerate and hence adjusted to maximize the neutrino production inside the jet. Now, using the same normalization of the proton spectrum required for this luminosity, we also calculate the luminosity in escaping UHECRs. Hence, it is essentially the same proton spectrum, with the same normalization. But, we assume at energies $\gtrsim0.1$ EeV they escape the source, because the observed SED does not allow for UHECR interactions inside the source and simultaneous explanation of quiescent state neutrino flux.

We present the methods of our one-zone modeling in Sec.~\ref{sec:model}. In Subsec.~\ref{subsec:sed} we present the results for leptohadronic emissions inside the jet. In Subsec.~\ref{subsec:uhecr} we show the contribution of line-of-sight resolved cosmogenic $\gamma$-ray contribution to the SED and the subsequent constraints on the EGMF strength. We discuss our results and draw our conclusions in Sec.~\ref{sec:discussions}.

%We describe the theoretical framework for our calculations and present our multiwavelength SED fits in Sec.~\ref{sec:results}. We discuss the implications of our finding on cosmic ray acceleration and draw our conclusions in Sec.~\ref{sec:discussions}. 

%--------------------------------------------------------------------

%--------------------------------------------------------------------

\section{\label{sec:model}Radiative Modeling}

%--------------------------------------------------------------------

%In our one-zone model, 
We consider the emission region in the jet 
%is considered 
to be a spherical blob of radius $R'$, consisting of a relativistic plasma of electrons and protons moving through a uniform magnetic field $B$. The bulk Lorentz factor of the jet is $\Gamma$ and the doppler factor is given by $\delta_D = [\Gamma(1-\beta\cos\theta)]^{-1}$, where $\beta c$ is the velocity of the emitting plasma and $\theta$ is the viewing angle. For $\theta\lesssim1/\Gamma$, 
%we can make the approximation 
$\Gamma\approx\delta_D$. We inject electrons in the blob with a spectrum
%assume a log-parabola spectrum for the electron injection, given by
%
\begin{align}
    Q'_e(\gamma'_e) = A_e(\gamma'_e/\gamma_0)^{-\alpha - \beta\log_{10}(\gamma'_e/\gamma_0)} \text{ \ \ for \  }  \gamma'_{e,{\rm min}} <\gamma'_e<\gamma'_{e, \rm{max}}
\end{align}
to fit the observed broadband SED. The normalization of the spectrum $A_e$ depends on the luminosity of injected electrons, and $\gamma_0m_ec^2$ is a reference energy fixed at 500 MeV. A quasi-steady state is reached when the injection is balanced by radiative cooling and/or escape. Empirically, the steady state electron density distribution is given as $N'_e(\gamma'_e)= Q'_e(\gamma'_e)t'_e$, 
%\src{Shouldn't it be $Q^\prime_e(\gamma^\prime_e)t^\prime_e$? Similarly in equations inline and below, time is in the jet frame and should be primed, no?} \textbf{saikat: Yes, it is a typo, I have corrected it}.
where $t'_e=$min\{$t'_{\rm cool}$, $t'_{\rm esc}$\}. We consider the escape timescale $t'_{\rm esc}\simeq t'_{\rm dyn}= 2R'/c$. The radiative cooling timescale is given as
\begin{align}
    t'_{\rm cool} = \dfrac{3m_ec}{4(u'_B+\kappa_{\rm KN}u'_{\rm ph})\sigma_T\gamma'_e}
\end{align}
Here $u'_B=B^2/8\pi$ is the energy density in magnetic field, $u'_{\rm ph}$ is the energy density of soft photons, $\sigma_T$ is the Thomson scattering cross-section, and $\kappa_{\rm KN}$ accounts for the suppression of IC emission due to the Klein-Nishina effect. We use the open-source code \textsc{GAMERA} to solve the transport equation for obtaining the injection spectrum at time $t'$ \citep{Hahn:2015hhw},
\begin{align}
    \dfrac{\partial N'_e}{\partial t} = Q'_e(\gamma'_e,t') - \dfrac{\partial}{\partial \gamma'_e}(bN'_e) - \dfrac{N'_e}{t_{\rm esc}}
\end{align}
where $b=b(\gamma'_e,t')$ is the energy loss rate of electrons. 

The steady-state electron spectrum yields the SYN and SSC emission. In addition, we consider an external photon field, which is Compton upscattered by the same %population of relativistic 
electrons. 
%This external photon field is indispensable for explaining the multiwavelength spectrum and also neutrino events from the source. 
It is considered to be a blackbody 
%spectrum parametrized by the 
with temperature $T'$ and energy density 
%($u'_{\rm ext}$) in the comoving jet frame. We have, 
$u'_{\rm ext} = (4/3)\Gamma^2 u_{\rm ext}$ in the jet frame, where the energy density in the AGN frame is $u_{\rm ext}=\eta_{\rm ext}L_{\rm disk}/4\pi R^2_{\rm ext}c$
%\begin{align}
%    u_{\rm ext}=\eta_{\rm ext}L_{\rm disk}/4\pi R^2_{\rm ext}c \label{eqn:ext}
%\end{align}
%
and $\eta_{\rm ext}$ is the fraction of the disk luminostiy. Here $R_{\rm ext}$ is the radius of the region containing the external photons. The emission blob is assumed to be at this distance along the axis of the jet.
%Assuming the blob is at a distance, along the axis of the jet, comparable to $R_{\rm ext}$, 
These photons can enter the relativistic jet and become doppler boosted in the comoving frame.

The steady state proton injection spectrum is given by a power-law $N'_p(\gamma'_p) = A_p\gamma_p'^{-\alpha}$. 
%
%\begin{equation}
%    Q'_p(\gamma'_p) = A_p\gamma_p'^{-\alpha}
%\end{equation}
%
The main energy loss processes of the protons are pion production ($p\gamma \rightarrow p+\pi^0$ or $n+\pi^+$) and BH process ($p\gamma\rightarrow p+e^+e^-$). The seed photons are the leptonic emission and external photons. The charged pions decay to produce neutrinos. The timescale of these interactions can be expressed as follows
\begin{equation}
\dfrac{1}{t'_{p{\rm \gamma}}} = \dfrac{c}{2\gamma_p'^2}\int_{\epsilon_{\rm th}/2\gamma_p}^\infty d\epsilon_\gamma'\dfrac{n(\epsilon'_\gamma)}{\epsilon_\gamma'^2}\int_{\epsilon_{th}}^{2\epsilon\gamma_p}d\epsilon_r \sigma(\epsilon_r) K(\epsilon_r)\epsilon_r \label{eqn:efficiency_ph}
\end{equation}
where $\sigma(\epsilon_r)$ and $K(\epsilon_r)$ are the cross-section and inelasticity respectively of photopion production or BH pair production as a function of photon energy $\epsilon_r$ in the proton rest frame. $n(\epsilon_\gamma')$ is the target photon number density \citep{Stecker:1968uc, Chodorowski_92, Mucke_00}. The interaction timescale of protons inside the jet is many orders of magnitude higher than the dynamical timescale, below tens of PeV energies. Hence, in order to increase the efficiency of $p\gamma$ interactions required for appreciable neutrino production, it is compelling to ignore their escape if the Eddington luminosity budget is maintained. Otherwise, the production of same neutrino flux will require higher kinetic power in protons, if the escape rate is comparable to $p\gamma$ interaction rate. Hence, an escape timescale higher than the $p\gamma$ interaction timescale is assumed. The normalization $A_p$ of the proton spectrum is calculated from the luminosity requirement arising from the in-jet hadronic contribution to the SED.

The spectrum of $\gamma$ rays from the decay of neutral pions and the spectrum of $\mathrm{e^+e^-}$ due to BH process are calculated using the parametrization by \citet{Kelner:2008ke}. When calculating the pion decay gamma-rays and electron spectra, the input proton spectrum is weighted by the rate of the corresponding process, for eg., in case of electron spectrum from Bethe-Heitler process, we inject $N'_p(\gamma'_p)*R_{\rm BH}/R_{\rm tot}$, where $R_{\rm BH}$ is the Bethe-Heitler interaction rate (calculated using Eqn.~4). $R_{\rm tot}$ is the total interaction rate considering photopion production and Bethe-Heitler pair production. The high-energy $\gamma$ rays are absorbed by $\gamma\gamma\to e^\pm$ pair production with the leptonic radiation and also with the external blackbody radiation, leading to the attenuation of TeV $\gamma$-rays. The escaping $\gamma$-ray flux is given as
\begin{equation}
    Q'_{\gamma, \rm esc}(\epsilon_\gamma') = Q'_{\gamma,\pi}(\epsilon'_\gamma) \bigg(\dfrac{1-\exp(-\tau_{\gamma\gamma})}{\tau_{\gamma\gamma}}\bigg)
\end{equation}
We calculate $\tau_{\gamma\gamma}$ using the formalism given by \citet{Gould_1967} to calculate the absorption probability per unit path length for an isotropic photon field,
\begin{align}
    l^{-1}_{\gamma\gamma}(\epsilon'_\gamma) = \dfrac{1}{2} \int\int n(\epsilon'_k) \sigma_{\gamma\gamma}(\epsilon'_\gamma, \epsilon'_k, \theta) (1-\cos\theta) \sin\theta d\theta d\epsilon'_k
\end{align}
where $\sigma_{\gamma\gamma}$ is the full pair-production cross-section and the $n(\epsilon'_k)$ is the combined density of soft photons and external radiation.

The high-energy electrons and positrons produced in $\gamma\gamma$ pair production ($Q'_{e,\gamma\gamma}$), charged pion decay ($Q'_{e,\pi}$), and BH process ($Q'_{e,\rm BH}$) can initiate cascade radiation from the jet.
%in the blob emission region. 
We solve the steady state spectrum of secondary electrons $N'_{e,s}(\gamma_e)$ in the 
%comoving 
jet frame using the analytical approach of \cite{Boettcher:2013wxa}, including $Q'_{e,\rm BH}$ in the source term and the escape term to be the same as primary electrons. In a synchrotron-dominated cascade, emission from secondary electrons is given by
\begin{align}
    Q'_s(\epsilon'_s) = A_0 \epsilon'^{-3/2}_s \int_{1}^{\infty} d\gamma'_eN'_{e,s}(\gamma'_e)\gamma'^{-2/3}_e e^{-\epsilon'_s/b\gamma'^2_e}
\end{align}
with 
%the normalization constant given as 
$A_0 =c\sigma_TB'^2/[6\pi m_e c^2 \Gamma(4/3)b^{4/3}]$ being a normalization constant, 
%
%\begin{align}
%    A_0 = \dfrac{c\sigma_TB'^2}{6\pi m_e c^2 \Gamma(4/3)b^{4/3}}
%\end{align}
%
where $b=B'/B_{\rm crit}$ and $B_{\rm crit} = 4.4\times 10^{13}$ G. 

Our results in Subsec.~\ref{subsec:sed} suggests that the proton spectrum inside the source is required to be cut off beyond a specific energy to explain the multiwavelength SED. The resulting neutrino flux is thus limited by this value of $E'_{p, \rm max}$. We model the protons to escape beyond this energy if accelerated inside their source. An energy-independent escape timescale of the order of $\sim R/c$ is sufficient for escape to dominate over photohadronic interactions inside the jet. However, considering a diffusion faster than $\propto E^1$ leads to negligible interaction efficiency beyond tens of PeV energies, in the comoving jet frame. This is reasonable when a quasi-ballistic propagation is assumed instead of diffusive propagation inside the jet emission region.

The resulting muon neutrino flux from $p\gamma$ interactions is calculated as
\begin{align}
    E_\nu^2J_\nu = \dfrac{1}{3} \dfrac{V'\delta_D^2\Gamma^2}{4\pi d_L^2}E'^2_\nu Q'_{\nu, p\gamma}
\end{align}
where the factor 1/3 corresponds to neutrino oscillation and $Q'_{\nu, p\gamma}$ is the total electron and muon neutrino flux from charged pion decay in the comoving frame.

%--------------------------------------------------------------------

\section{\label{sec:results}Results}
\subsection{\label{subsec:sed}Leptohadronic emission inside the jet}

%--------------------------------------------------------------------

%--------------------------------------------------------------------
%
%
%\begin{figure}
%\centering
%\includegraphics[width=0.48\textwidth]{e_secondary.pdf}
%\caption{\small{The spectrum of secondary electrons from Bethe-Heitler process (dotted line), charged pion decay (dashed-dotted line), and the pair production due to collision of $\pi^0$-decay $\gamma$-rays and the SED photons, in the comoving jet frame for the low-state.}}
%\label{fig:elec}
%\end{figure}
%
%
%--------------------------------------------------------------------
%
%
\begin{figure*}
\centering
\includegraphics[width=0.49\textwidth]{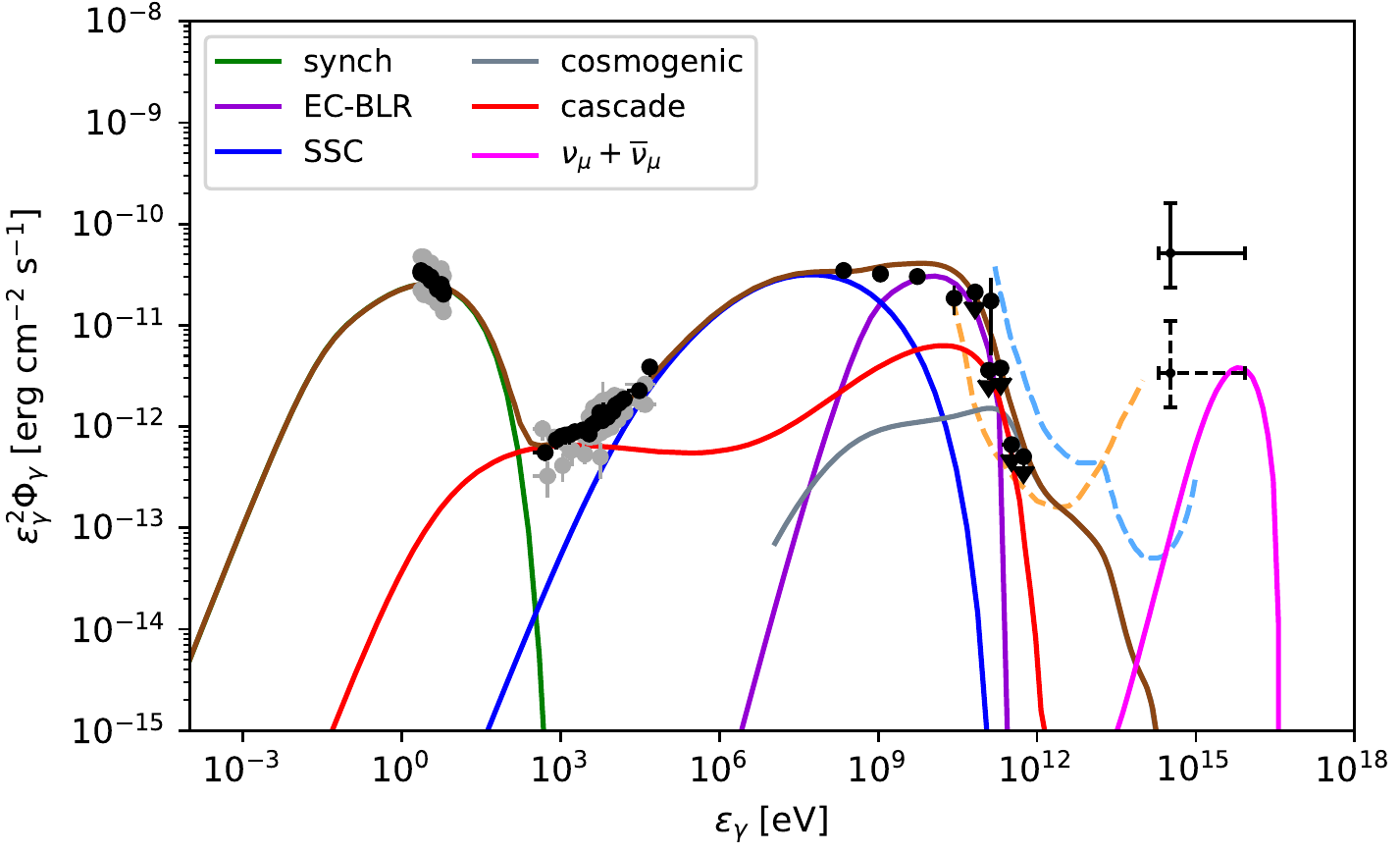}
\includegraphics[width=0.49\textwidth]{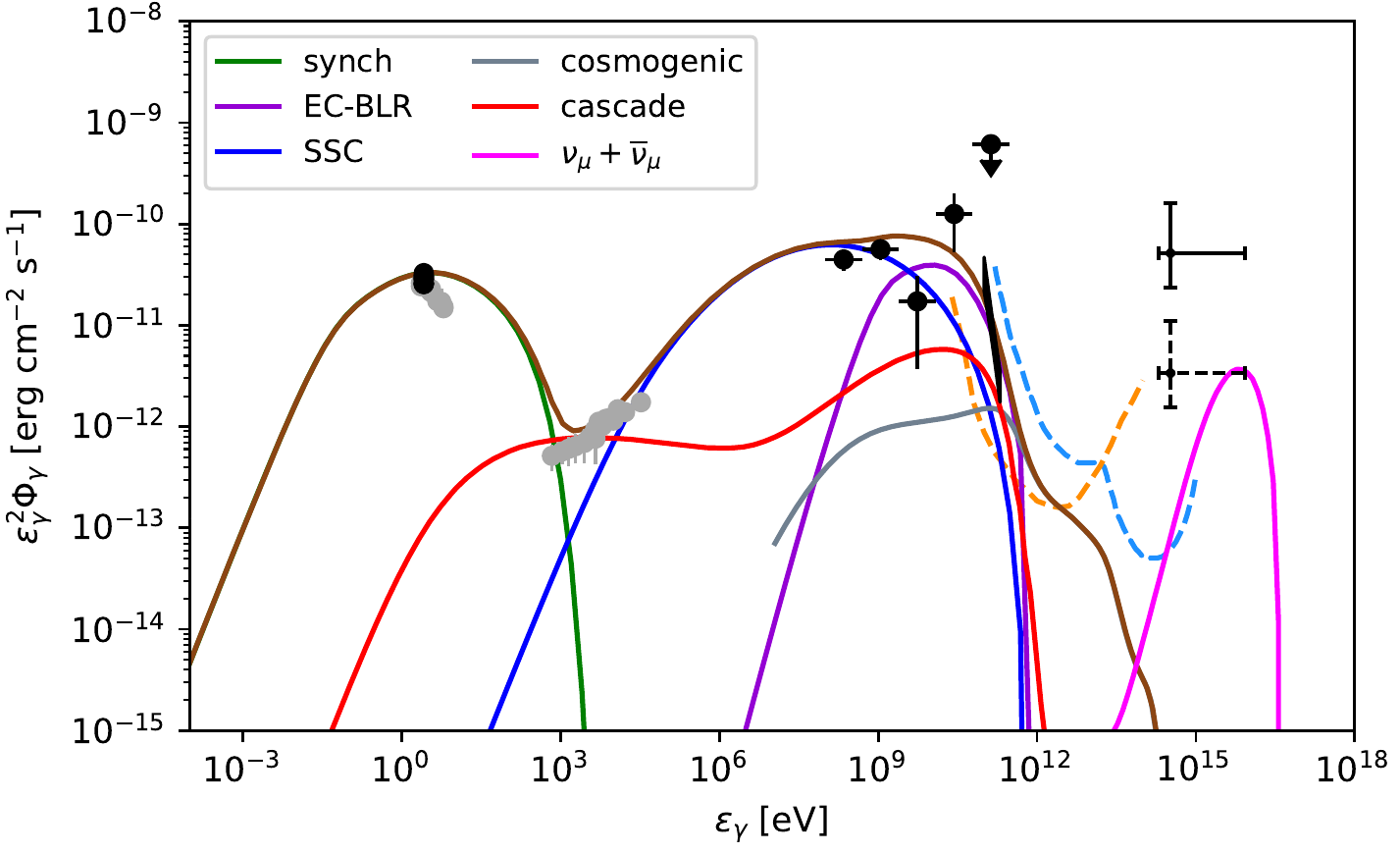}
\caption{\small{Multiwavelength SED of TXS 0506+056 during the 2017-2019 campaign. \textit{Left:} Low-state (average) flux during the observation period. Black data points show the Fermi-LAT average spectrum and MAGIC upper limits, and Swift and NuSTAR spectra on 2018 October 16. Gray data points show the whole range of optical-to-X-ray spectra. \textit{Right:} High-state observation in VHE $\gamma$-rays during 2018 December. Black data points show the Fermi-LAT, MAGIC, and ASAS-SN contemporaneous spectra. Gray data points show the nearest observations in radio, optical, UV, and X-rays on 2018 December 8. The green, blue, and purple curves corresponds to SYN, SSC, and EC-BLR emission. The red curve is the cascade emission from secondary electrons. The magenta curve is the predicted neutrino flux in both cases. The black solid and dashed PeV data points represent the IceCube flux upper limit for one detection in 0.5 yrs and 7.5 yrs respectively for the 2017 detection event. The grey line is the cosmogenic $\gamma$-ray flux from extragalactic propagation of UHE protons. The orange and blue dashed lines are CTA and LHAASO point source sensitivity. See text for details.}}
\label{fig:low}
\end{figure*}
%
%
%--------------------------------------------------------------------

During the multiwavelength campaign, the source was monitored using the Swift-XRT, Swift-UVOT, and NuSTAR, maximizing the simultaneity of observation with the MAGIC telescope \citep{MAGIC:2022gyf}. From November 2017 to February 2019, a total of $\sim79$ hrs of good-quality data was collected by MAGIC. During most of this period ($\sim74$ hrs), the source was found to be in a low-state with average photon flux $F$($>90$ GeV)$=(2.7\pm 2.1)\times10^{-11}$ erg cm$^{-2}$ s$^{-1}$. 
%Publicly available 
Fermi-LAT data was also used in the spectral analysis. For the flaring state of December 2018, only simultaneous data is obtained by Fermi-LAT and MAGIC in the GeV
%high-energy 
and VHE $\gamma$-rays, and by ASAS-SN in the optical. 
%waveband. 
The integral photon flux observed by MAGIC rose by an order of magnitude compared to the low state. The most significant variability was observed in the GeV
%Fermi-LAT $\gamma$-ray 
band, while the X-ray variability was found to be at a lower level. The radio, optical and UV bands showed moderate variability \citep{MAGIC:2022gyf}.

%The redshift of the blazar is $z=0.3365$, the corresponding to a 
%The relativistic plasma is described fully by the bulk Lorentz factor $\Gamma$, the radius $R'$ and the magnetic field $B'$. Assuming the line-of-sight to be within the jet opening angle, the bulk Lorentz factor $\Gamma\approx\delta_D$, which is considered to be fixed for the low-state and high-state in our analysis. 

The $\gamma$-ray variability timescale of TXS~0506+056 observed in October 2017 was shown to be $t_{\rm var}\leq10^5$ s \citep{Keivani:2018rnh}. The $\gamma$-ray flare of December 2018 was found to be very similar.
%to that observed in October 2017, in terms of flux level and day scale variability. 
The size of the emission region inferred from the variability is $R'\lesssim \delta_Dct_{\rm var}/(1+z) \simeq 6.75\times 10^{16}(\delta_D/30)(t_{\rm var}/10^5 \text{s})$ cm. 
%In our modeling 
We assume the radius of the emission region to be $10^{16}$ cm and $\Gamma\approx\delta_D$ during both the high- and low-states. The value of the magnetic field was fine-tuned by fitting the optical and gamma-ray data. It is also assumed to be the same in the two states, $B'=0.28$ G. The muons and pions produced in hadronic interactions do not suffer significant energy losses before decaying, for this value of $B'$.

The luminosity distance of TXS~0506+056 is $d_L\approx 1837$ Mpc, with a redshift of $z=0.3365$. The total kinetic power of the jet in the AGN frame is 
%evaluated 
calculated as $L_{\rm kin} = L_e + L_p + L_B = \pi R'^2\Gamma^2c(u'_e+u'_p+u'_B)$, where $u'_e$, $u'_p$, and $u'_B$ are the energy densities of electrons, protons, and magnetic field respectively. The maximum electron energy changes from $\gamma'_{e,{\rm max}}=2\times10^{4}$ in the low-state to $\gamma'_{e, {\rm max}}=5\times10^{4}$ in the high-state to account for the spectral variability. We vary the maximum proton energy ($E'_{p, {\rm max}}$) in the comoving jet frame over a wide range to find the best-fit value of 6.3 PeV, fixed for both the low- and high-states. 
%The spectrum of secondary electrons from various processes and $\gamma$-rays from $\pi^0$ decay is shown in Fig.~\ref{fig:elec} for the low-state. 
The cascade emission from the steady-state secondary electron spectrum $N'_{e,s}$ is shown by the red lines in Fig.~\ref{fig:low}. The low energy peak of the cascade emission originates from the secondary emission of $e^\pm$ pair produced in BH process, which is severely constrained by the X-ray data. As a result the pion decay cascade at higher energies is also limited and the contribution to the high-energy peak is not significant.

%--------------------------------------------------------------------
%
%
\begin{table}
\centering
\caption{\label{tab:mwl_fit}Model parameters for the multiwavelength SED, indicating the electron and proton luminosites in the AGN rest frame
%\src{What is the value of $\alpha$ for protons?}
%\textbf{saikat: we have kept it same for both protons and electrons, $\alpha=2$}
}

\begin{tabular}{ccc}
\hline
\textbf{Parameters} & \textbf{Low State} &  \textbf{High State}\\
\hline
$\delta_D$ & 28 & ''\\
$B'$ [G] & 0.28& ''\\
$R'$ [cm] & $10^{16}$ & ''\\
$u'_{\rm ext}$ [erg/cm$^3$] & 0.01 & ''\\
$T'$ [K] & $2\times10^5$& ''\\
$\alpha$ ($e$/$p$ spectral index) & 2.0 & ''\\
$\beta$ (log parabola index)  & 0.3 & ''\\
$E_0$ [MeV] & 500& ''\\
\hline
$E'_{\rm e,min}$ [GeV] & 0.20 & 0.25\\
$E'_{\rm e,max}$ [GeV] & 10 & 25\\
$L_e$ [erg/s] & $5.8\times10^{44}$ & $7.6\times10^{44}$\\
$E'_{\rm p, min}$ [GeV] & 10 & ''\\
$E'_{\rm p,max}$ [PeV] & 6.3 & '' \\
$L_p$ [erg/s] & $1.6\times10^{48}$ & '' \\
\hline
\end{tabular}
\end{table}
%
%
%--------------------------------------------------------------------

We obtain $T'=2\times10^5$ K and $u'_{\rm ext} = 0.01$ erg/cm$^3$ for the external photon field from fitting the SED, 
%Apart from being 
which is the most important target of $p\gamma$ interaction for neutrino production 
%they are also seed photons 
and for IC scattering, crucial
%and is the key 
to explain the VHE spectrum. It is also vital for $\gamma\gamma$ absorption in the jet, beyond a few hundreds of GeV. 
%Considering 
For a typical disk luminosity $L_{\rm disk}\approx 10^{46}$ erg/s and the scattered disk emission to be a fraction $\eta_{\rm ext}\sim 0.01$ of the disk photon energy density, $R_{\rm ext}$ comes out to be a few times $10^{18}$ cm. 

The VHE flare of December 2018 does not have simultaneous observation at lower energies. Thus the constraints on the theoretical model are moderate. Nevertheless, to reduce the uncertainties, only the electron primary distribution and its corresponding luminosity is changed with respect to the low state. The parameter values used in the modeling are given in Tab.~\ref{tab:mwl_fit}.

%First, 
We fit the low-state spectrum first and optimize the parameters $\delta_D$, $B'$, and spectral indices, using the leptonic emission only. The SYN spectrum peaks at the optical band and a log-parabola injection spectrum of electrons well explain the data. 
%The SYN and SSC flux is doppler boosted by a factor of $\delta_D^4$ in the observer frame. The EC-BLR flux is also boosted by a factor of $\delta_D^6/\Gamma^2$. 
The hadronic component is then added and the power and maximum proton energy is varied to fit the X-ray data with BH cascade. The VHE photon flux upper limits constrain the contribution from the pion-decay cascade. We also consider the absorption of VHE $\gamma$-rays in the extragalactic background light (EBL) using the Gilmore et al. model \citep{Gilmore_2012}. It can be seen from the left panel of Fig.~\ref{fig:low}, that the neutrino flux is comparable to the 7.5-year averaged flux prediction from this source by IceCube. We find the neutrino flux to be roughly unchanged in modeling the low- and high-state SEDs obtained by the MAGIC multiwavelength campaign. This corresponds to a flux that produces on average one neutrino detection like IC-170922A over a period of 7.5 years and explains the non-observation of neutrino events during the December 2018 flare in the MAGIC waveband. A further increase in the neutrino flux leads to the violation of the X-ray data.

%The maximum energy of protons that undergo photohadronic interactions in the AGN frame is then $E_{p, \rm max}=\Gamma E'_{p, \rm max}$, which corresponds to $\sim$0.17 EeV in the AGN frame. 

\subsection{\label{subsec:uhecr}Cosmogenic $\gamma$-rays from UHECRs}
The maximum proton energy for photohadronic interactions in the AGN frame is $E_{p, \rm max}=\Gamma E'_{p, \rm max} \approx 0.17$ EeV from modeling MWL SEDs of TXS 0506+056. The proton spectrum has to be cut off at $\mathcal{O}\sim 0.1$ EeV inside the jet to satisfy constraints from X-ray data, and to simultaneously produce neutrinos with a flux in the PeV range inferred from detection of one event in 7.5 yr of IceCube operation.
%explain the astrophysical neutrino production. 
According to the Hillas condition, the maximum acceleration energy $E^{\rm acc}_{p, \rm max} \sim 2\beta c ZeBR\Gamma$, where the bulk Lorentz factor $\Gamma$ takes into account the frame transformation from comoving jet frame to AGN frame. The gyration radius of $10^{20}$ eV protons from this simplistic expression may be calculated as $r_{\rm L}\approx 2.13\times 10^{16}$ cm, which is comparable to the blob radius in our modeling. Thus protons of energy higher than 0.17 EeV is possible to be produced in the same blob for the magnetic field and the length scale considered. The $p\gamma$ opacity in jet is higher than 1e-3 for protons with energy $>6\times10^{17}$ eV \citep[see eg.][]{Das:2021cdf}. The jet is opaque to photons beyond $\sim 1$ TeV, therefore if UHE protons interact inside the jet, additional gamma-rays from $\pi^0$, $\pi^\pm$ cascade would contribute at lower energies, violating the X-ray data. Hence, we assume that UHE protons beyond this energy escape the jet. A faster escape at higher rigidities has been parametrized in earlier studies. For eg., see Eqn.~20 in \cite{Harari:2013pea} and applied to the case of inside the source in \cite{Muzio:2021zud}. Thus, it is inherently the same population of protons, which interacts below $\sim0.1$ EeV and escapes at higher energy. This assumption allows us to use the same normalization of the proton spectrum inside and outside the jet, with the same injection spectral index.

Assuming that protons escape efficiently beyond $\sim0.1$ EeV
and up to $10^{20}$ eV as UHECRs, we calculate the cosmogenic $\gamma$-ray spectrum resulting from their propagation in the EGMF and interactions with the CMB and EBL photons. 
%of UHECR protons. We consider the maximum acceleration energy of protons can reach $\sim100$ EeV. 
%These protons can interact with the cosmic microwave background (CMB) and EBL to produce secondary $\mathrm{e^\pm}$ and $\gamma$-rays. In particular, 
%the resonant photopion production of UHE protons, near GZK energy, with CMB can efficiently produce pion decay $\gamma$-rays. These high-energy secondaries can 
Secondaries from these interactions initiate electromagnetic cascade in the extragalactic medium, undergoing synchrotron radiation in the EGMF, pair production with EBL, inverse IC scattering of background photons, etc. The resulting spectrum is shown by the grey line in Fig.~\ref{fig:low}. Note that the cascade is sufficiently developed for $E_{p} \gtrsim 40$ EeV, the GZK energy. Hence, the spectrum depends more on the propagation effects than on the source parameters. 
%CTA is an upcoming imaging telescope that 
The upcoming $\gamma$-ray detector Cherenkov Telescope Array (CTA) will detect $\gamma$ rays
%capable of detecting $\gamma$-rays from 
in the range from 20 GeV up to several hundred TeV, with unprecedented sensitivity. The orange dashed curve in Fig.~\ref{fig:low} shows the differential point source sensitivity of CTA, assuming 50h observation time and pointing to 20 degrees zenith \citep{Gueta:2021vO}. CTA observation of a hard and non-variable multi-TeV $\gamma$-ray spectrum will indicate the presence of UHECR acceleration inside this source. The blue dashed line is the LHAASO 1-yr sensitivity to Crab-like $\gamma$-ray point sources \citep{Vernetto:2016gro}.

\begin{figure}
\includegraphics[width=0.45\textwidth]{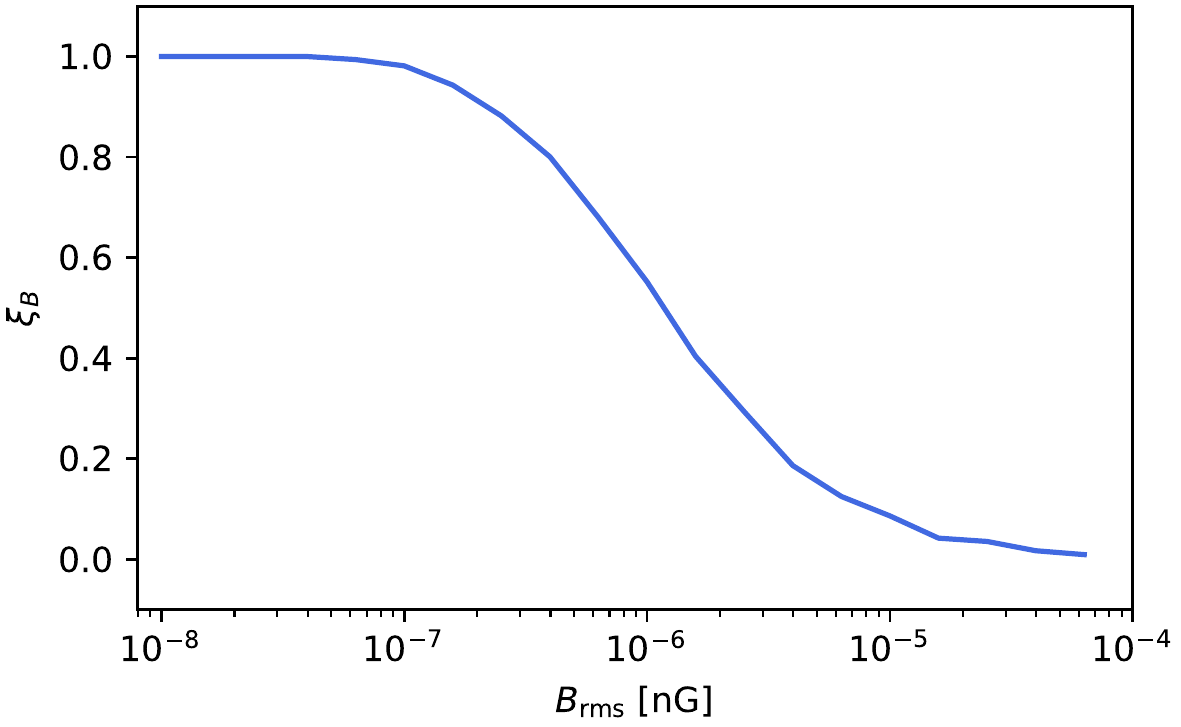}
\caption{\small{Survival fraction of UHECRs along the line-of-sight, i.e., within $0.1^\circ$ of the initial direction of propagation, as a function of the rms strength of EGMF.}}
\label{fig:brms}
\end{figure}

%Now, UHECRs are deflected by the EGMF during propagation. 
We define the line-of-sight resolved component of the cosmogenic $\gamma$-ray spectrum as the fraction of UHECRs ($\xi_B$) that survives within $1^\circ$ of the initial propagation direction, from the source to the observer, after deflection in the EGMF. For neutrinos of energy $\sim30$ TeV, the angular resolution of IceCube for track-like events is $\sim0.5^\circ$. Whereas, Fermi-LAT has a resolution of $3.5^\circ$ to photons of energy $<100$ MeV, and $\sim 0.15^\circ$ 
%for photons 
beyond 10 GeV. We use 
%the public simultaion tool 
CRPropa-3 to propagate UHECRs in the extragalactic space \citep{AlvesBatista:2016vpy}. $\xi_B$ is calculated from the arrival direction of protons in a 3D simulation on the surface of a sphere of radius 100 kpc, centered at the observer. We propagate a $E^{-2}$ proton spectrum in the energy range between 0.1 and 100 EeV 
%and consider 
in a random turbulent magnetic field with a Kolmogorov power spectrum. The 
%turbulence correlation 
coherence length of the field is adjusted to $100$ kpc. Thus, the secondary flux for a given luminosity is multiplied by $\xi_B$ to obtain the line-of-sight component at the position of the observer \citep{Das:2019gtu}. The survival fraction $\xi_B$ as a function of rms field strength is shown in Fig.~\ref{fig:brms}. 

The value of $B_{\rm rms}$ can be constrained from the required luminosity in cosmogenic $\gamma$-rays by the following expression, under isotropic approximation
\begin{equation}
 \dfrac{L_{{\rm UHE}p}}{4\pi d_L^2} = \dfrac{1}{\xi_B f_{\gamma,p}} \int_{\epsilon_{\gamma,\rm min}}^{\epsilon_{\gamma,\rm max}} \epsilon_\gamma\dfrac{dn}{d\epsilon_\gamma dAdt}d\epsilon_\gamma \label{eqn:lum}
\end{equation}
where ${dn}/{d\epsilon_\gamma dAdt}$ is the differential flux of cosmogenic $\gamma$ rays constrained by the SED. We calculate the electromagnetic cascade using the external code DINT integrated with CRPropa-3 \citep{Heiter_18}. The factor $f_{\gamma,p}$ takes into account the fraction of injected UHECR power that goes into cosmogenic $\gamma$ rays and is fairly constant with the variation of $B_{\rm rms}$. For TXS~0506+056
%a cosmic ray source at $z=0.3365$, 
we find $f_{\gamma,p}\approx0.156$. The integrated flux of cosmogenic photons along the line-of-sight of the observer, allowed by the observed SED is $\sim1\times10^{-11}$ erg cm$^{-2}$ s$^{-1}$, as found from Fig.~\ref{fig:low}. The luminosity of protons interacting inside the jet is found to be $L_p = 1.6\times10^{48}$ erg/s (cf.\ Tab.~\ref{tab:mwl_fit}). Using the same normalization for the escaping proton spectrum beyond 0.1 EeV,
%for the proton spectrum, 
the luminosity in UHECR protons, i.e., between $0.1-100$ EeV is $L_{{\rm UHE}p}\approx8\times10^{47}$ erg/s.  This implies from Eq.~(\ref{eqn:lum}), for the allowed flux of cosmogenic $\gamma$ rays, 
%the survival fraction 
$\xi_B\lesssim 0.05$. This indicates an EGMF with RMS value higher than few times $10^{-5}$ nG as seen from Fig.~\ref{fig:brms}. There may be no cosmogenic component along the line of sight for magnetic field strength much larger than this, otherwise, the luminosity budget is violated. It is to be noted, that this result is an order of magnitude estimate. The precise value is sensitive to the angular resolution to detect high energy $\gamma$-rays, the coherence length of EGMF, the angular spread of jet emission, and the numerical precision of cascade calculation.

%----------------------------------------------------------------- 
%
%
\begin{figure}
\includegraphics[width=0.48\textwidth]{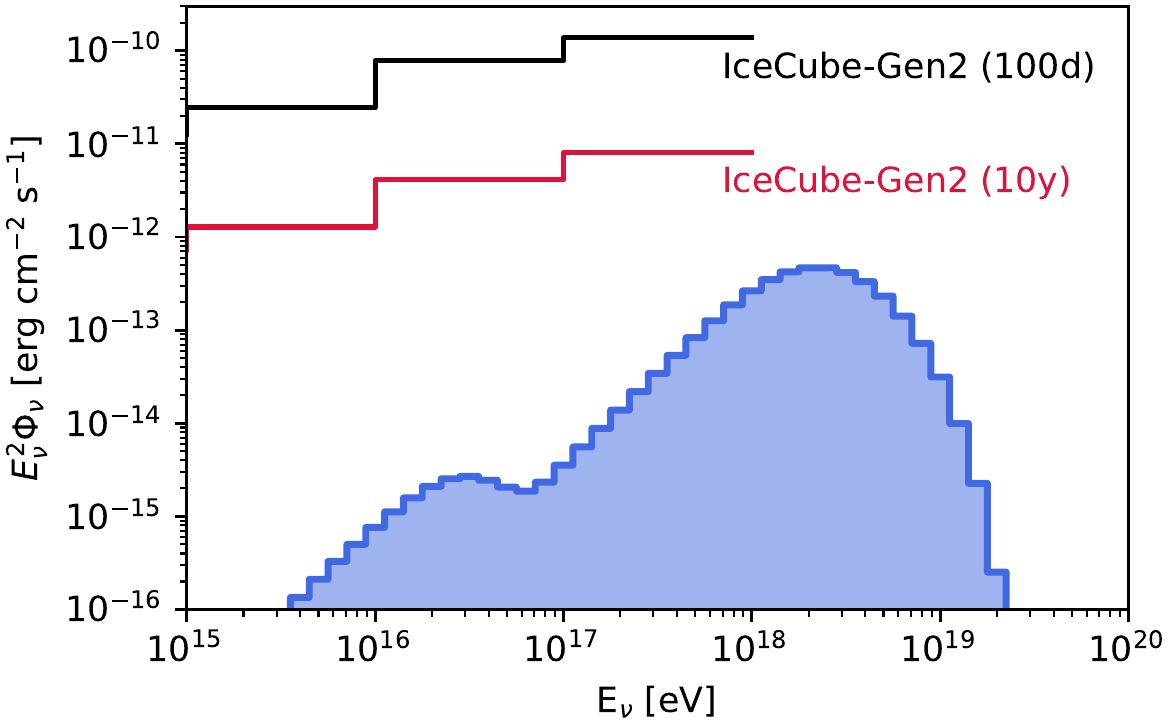}
\caption{\small{
%The resulting 
All-flavor cosmogenic neutrino flux from TXS 0506+056 due to UHECR propagation 
%and that resolved 
along the line of sight, i.e., using the same normalization as the cosmogenic $\gamma$-ray spectrum in Fig.~\ref{fig:low}. The black and red curves correspond to 100 days and 10 years of observations of muon neutrino fluxes by IceCube-Gen2 and indicate the sensitivity for neutrino flares and the time-averaged neutrino emission, respectively.}}
\label{fig:neu}
\end{figure}
%
%
%----------------------------------------------------------------- 

We calculate the flux of cosmogenic neutrinos produced simultaneously during the propagation of UHECRs and use the same normalization as obtained for the allowed flux of cosmogenic $\gamma$-ray spectrum. The cosmogenic neutrino spectrum peaks at $2.8\times10^{18}$ eV with peak flux $\sim 4.5\times10^{-13}$ erg cm$^{-2}$ s$^{-1}$, shown in Fig.~\ref{fig:neu}. The IceCube-Gen2 detector will be capable of detecting neutrinos from TeV to EeV energies, with sensitivity five times larger than the currently operating IceCube experiment. We also show the $5\sigma$ sensitivity for detection of muon neutrino flux from TXS 0506+056, using the IceCube-Gen2 detector \citep{IceCube-Gen2:2020qha}. The black and red lines correspond to 100 days and 10 years of observation and indicate the sensitivity for neutrino flares and the time-averaged neutrino emission, respectively. The neutrino flux is lower than the detection threshold, however. Thus the $\gamma$-rays provide more stringent limits on UHECR acceleration in TXS~0506+056.

%----------------------------------------------------------------- 
%
%
\section{\label{sec:discussions}Summary and Conclusions}

%Recent analysis has found that 10 out of 19 IceCube hotspots in the southern sky have a positional correlation with blazars \citep{Buson:2022fyf}. Their study indicates blazars accelerate cosmic rays at least up to PeV energies, and possibly up to EeV energies. However, the contribution of $\gamma$-ray blazars to diffuse high-energy neutrino flux remains small, also indicating a weak correlation between GeV $\gamma$-ray emission and neutrino production in blazars. Detection of X-ray flare will be crucial to test the one-zone photohadronic scenario, and imply that blazar neutrino emission is most prominent during the flaring episodes.

The flux variability observed by MAGIC in December 2018 was very similar to that seen in 2017. No neutrino event was detected in 2018 however, in contrast to the 2017 flare. In our one-zone modeling of the multiwavelength data from Nov 2017 - Feb 2019, we see that an increased $\gamma$-ray activity does not yield an increased neutrino flux, and the latter is comparable to the 7.5-yr flux prediction by IceCube, same as that obtained for the low state. Hence, the correlation between $\gamma$-ray and neutrino activity is subjective to the model undertaken. In our one-zone modeling we include three components, viz., the leptonic emission inside the jet, secondary emission due to hadronic cascade induced by protons ($E_p \lesssim 0.1$ EeV) and the line of sight resolved cosmogenic $\gamma$-rays due to propagation of UHECR protons ($E_p \gtrsim 0.1$ EeV) from the blazar to the Earth. The emission from the cascade is found to be sub-dominant in the GeV range but important in the X-ray and VHE bands. If the same physical process is responsible for the neutrino production, then the X-ray data constrains the neutrino flux to be consistent with IceCube prediction for one event in $\Delta T = 7.5$ yrs. Hence, an additional hidden sector must be invoked to explain a higher neutrino flux, for e.g., the neutrino flare in 2017 and $\Delta T = 0.5$~yr flux. This is consistent with no observation of $\gamma$-ray activity during the orphan neutrino flare during 2014-15.

We highlight our modeling as an important step in the course of study of TXS 0506+056, viz., (i) a neutrino flare (archival data) was observed in 2014-2015, but no gamma-ray activity, (ii) then an increased gamma-ray activity and simultaneous detection of a neutrino event in 2017 (iii) finally, no neutrino detection but increased gamma-ray activity in 2018. From the study of phase (iii), we show that indeed there is little correlation between gamma-ray and neutrino flares. \cite{Petropoulou:2019zqp} put an upper limit of $(0.4-2) \ \nu_\mu$ events in 10 years of IceCube operation, through multi-epoch modeling. They argue that the IC-170922A can be explained as an upward fluctuation from the average neutrino rate expected from the source, but in strong tension with the 2014–2015 neutrino flare. \cite{Rodrigues:2018tku} shows only $2-5$ neutrino events during the 2014-15 flare can be explained consistently with the X-ray constraints or high-energy $\gamma$-ray flux measured by Fermi-LAT. For all the cases presented in the two-zone model of \cite{Xue:2019txw}, the neutrino flux prediction is comparable to the 7.5-yr IceCube upper limit. In our study, the neutrino event rate is $\mathcal{N_{\nu_\mu+\overline{\nu}_\mu}}=1\times(\Delta T/7.5 \ \rm yrs)$. Hence, along the lines of the one-zone leptohadronic model adopted here, our results lead to similar conclusions using yet another epoch of the blazar and thus add to the literature.

The lack of simultaneous X-ray data in the high state is a drawback to the SED modeling, although the data points shown are that for the nearest observation on 2018 December 8. The parameters are varied minimally from the low state to account for this. Interestingly, our modeling does not predict any significant flare of the optical flux, where the SYN spectrum peaks, but the UV and soft X-ray fluxes are expected to change moderately. We note that a higher X-ray flux can allow for an increased neutrino flux, however. But explaining the $\Delta T=0.5$ yr neutrino flux remains difficult due to excess X-ray production. Many plausible alternatives exist in the literature, such as neutrino production near the supermassive black hole of the AGN, in accretion disk or corona \citep{Stecker_2013, IceCube:2021pgw}, or multiple emission zones with increased $\gamma\gamma$ opacity in the neutrino production zone \citep{Xue:2020kuw}, etc. Nevertheless, production of one neutrino event in $0.5$-yrs is achieved in \cite{Cerruti:2018tmc, Fraija:2020zjk}, etc., but the neutrino flux peak is shifted compared to the mean energy of the observed event.

The origin of external photons is a question of fundamental importance in the modeling of TXS-like blazars. In their modeling, MAGIC collaboration used an external field originating from the spine layer or the jet-sheath 
%was considered 
\citep{MAGIC:2018sak, MAGIC:2022gyf}. In our analysis, we consider it to originate from the BLR. It provides a substantial target for neutrino production by $p\gamma$ processes and also inverse-Compton scattering by electrons. For this to be true, the radius of the emission region must be smaller than the radius of the BLR region. In our analysis, $R_{\rm ext}$ is few times $\sim 10^{18}$ cm, which is large compared to usual estimates for the BLR region, $R_{\rm BLR}=10^{17} L_{\rm disk, 45}^{0.5}$ cm \citep[see Eqn.~4 in][]{Ghisellini:2008zp}. One possibility is that the blob lies at the edge or even outside the BLR region leading to a decrease in the effective BLR photon density \citep{Tavecchio:2008vq}. The typical energy of the photons in the AGN frame is $\epsilon_{\rm ext}\sim 3k_BT'/\Gamma \approx 17$ eV. This is comparable to that obtained in other studies \citep{Keivani:2018rnh} and can also be considered as scattered emission from the disk. The contribution from disk photon itself is negligible. We consider a log-parabola spectrum for the injection of electrons, to improve the fit to the observed SED. Often, other assumptions are also made in the literature, such as a broken power-law spectrum \cite{Xue:2019txw}.

In our modeling, for photohadronic interaction rate to dominate over escape, we need an escape timescale higher than $10^6 (R/c)$ at $\sim 1$ PeV inside the emission region, and even higher at lower energies, assuming an energy-independent escape. In a energy-dependent parametrization, the escape timescale can be expressed as $t_{\rm esc} \gg 10^6 (R/c) (E/10^3 \text{\ TeV})^{-1}$ for the proton energy range interacting inside the jet. In the one-zone model,  the efficient escape of UHECRs require a rigidity-dependent diffusion rate, for e.g., $D(E)\propto E^2$ at higher energies \citep{Globus:2007bi, Harari:2013pea, Muzio:2021zud}. As an alternative to the step function for the escape, as we had assumed, one may also assume a separate emission zone for acceleration of UHE protons with lower photohadronic opacity. We do not present the analytical estimates of such an astrophysical scenario in this paper, but assume a single proton population.
%, devoid of target photons or gas. 
In our analysis, the cosmogenic $\gamma$-ray spectrum remains fixed for both the low- and high-states. A change in the 
%internal proton 
primary proton distribution will not affect the cosmogenic flux significantly, because the spectrum is driven greatly by parameters guiding the extragalactic propagation. Since UHECRs are delayed in the EGMF, any observed variability in the VHE regime occurs, most likely, due to an increased activity inside the jet. The required luminosity in UHE protons can also be translated into a resulting flux of neutrinos at EeV energies. The cosmogenic neutrino flux predicted here, from constraints on $\gamma$-ray flux, is found to be lower than that in our earlier study \citep{Das:2021cdf}. Thus detection of cosmogenic neutrinos from TXS 0506+056 seems unlikely with the next generation upgrade of IceCube, leaving ground-based $\gamma$-ray detectors such as CTA to test UHECR signature in the SED of blazars.
%
%
%----------------------------------------------------------------- 

\begin{acknowledgements}
      S.D. thanks Konstancja Satalecka (MAGIC Collaboration) for correspondence regarding the multiwavelength SED data. The work of S.D. was supported by JSPS KAKENHI Grant Number 20H05852. Numerical computation in this work was carried out at the Yukawa Institute Computer Facility. S.R. was supported by a grant from NITheCS and the University of Johannesburg URC. We thank the anonymous referee for useful comments and suggestions.
\end{acknowledgements}

\bibliographystyle{aa} % style aa.bst
\bibliography{magic.bib}

\end{document}